\documentclass{appolb}
\usepackage{graphicx}

\usepackage{amsmath}

\def\vec#1{\ensuremath{\mathchoice
                     {\mbox{\boldmath$\displaystyle#1$}}
                     {\mbox{\boldmath$\textstyle#1$}}
                     {\mbox{\boldmath$\scriptstyle#1$}}
                     {\mbox{\boldmath$\scriptscriptstyle#1$}}}}
\begin{document}
\title{ A study of retardation models for phenomenological interactions in hadronic physics
}
\author{M. De Sanctis \footnote{mdesanctis@unal.edu.co}
\address{Universidad Nacional de Colombia, Bogot\'a, Colombia }
\\[3mm]
}

\maketitle
\begin{abstract}
A study of  retardation models for the hadronic quark interaction is performed 
starting from
a coordinate space calculation  that elaborates a classical electrodynamics procedure.
The possibility of constructing a corresponding quantum operator is critically analyzed
also performing some numerical matrix element calculations.
A comparison of the developed model  with the Feynman diagram interaction at tree-level 
is studied,
showing a substantial physical correspondence of the two models.
Possible applications and generalizations are discussed.

\end{abstract}

	
\section{Introduction}
The study of hadronic systems in terms of interacting constituent quarks still
represents an interesting field of investigation, 
due to the nonperturbative character of QCD and to the difficulties 
of the lattice calculations.

Among the different issues, we study in the present paper the problem of
\textit{retardation} in the quark strong interaction.
Some contents of this work also appear in Ref. \cite{arxret}, 
where the didactic aspects are emphasized. 
%
%
%
%
In the present work we focus our attention on \textit{heavy} $q ~ \bar q$ systems,
such as charmonium and bottomonium. 
The dynamics of these hadronic system,
composed by two equal mass particles, 
can be reproduced, in their center of mass (CM) reference frame, by means of a relativistic 
wave equation in which the constituent quark interaction is represented by
a specific coordinate space potential.
 
In this respect, we cite here a series of works developed by the author
in the framework of a three-dimensional reduction of a two-body Dirac equation
\cite{chromomds,localred,rednumb,relvar,scalint,effective,explor}.
The references to other
relativistic models for hadronic systems can be found particularly in Ref. \cite{localred}.

It has been possible to show that a good reproduction of the charmonium data
can be obtained by taking a \textit{vector} interaction term,
that is reminiscent of the underlying QCD dynamics, plus a scalar (or mass) term 
that reproduces the correct spin-orbit splittings  and can be related
to the scalar resonances of the hadronic spectrum \cite{rednumb,scalint}.
However, the main role in the interaction is played by the vector term,
that can be represented by a Coulombic potential, \textit{i. e.} given by the interchange of a massless particle, regularized by a chromo-electric charge distribution of the interacting quarks \cite{chromomds}.

The interaction vector potential is usually derived in the framework of a three-dimensional 
relativistic wave equation, that is obtained by \textit{reducing} the four-dimensional
Bethe-Salpeter equation \cite{sabe,bethes}.
This treatment, that was originally developed for the study of \textit{electromagnetically} 
bound systems, allows to add the retardation corrections by means of a cumbersome procedure 
\cite{itz}
that relies on the smallness of the coupling constant and, for this reason, 
is not straightforwardly
applicable in the case of strongly interacting particles. 

Retardation corrections together with
other terms of relativistic corrections
have been also added, perturbatively, to a standard Schr\"odin\-ger effective Hamiltonian
\cite{fermibreit,charmrc}.
But this technique is now superseded by fully relativistic calculations.

In the standard \textit{instantaneous} or \textit{unretarded} approach,
the bound system is studied by means of
 a relativistic three-dimensional equation
in which the interaction is represented by 
a potential operator that depends only on the interquark distance
$r$, completely ignoring the retardation effects of the interaction.
As it will be recalled in Subsect. \ref{unret}, this modelization
(studied in more detail in Ref. \cite{relvar})
 does not violates relativity
\textit{per se}. 
 Phenomenologically, 
it is able to reproduce accurately the charmonium data.

Nevertheless, it is worthwhile to investigate the possible effects related to retardation,
taking into account that retardation
is generally considered a specific characteristic of particle interactions in field theory
and could give nonnegligible contributions in strongly interacting systems.
%
\vskip 0.5 truecm

This last point represents  the main motivation of the present work.
In other words, we show that retardation (or \textit{relative time effect}), which is usually
lost when  reducing the four-dimensional Bethe-Salpeter equation \cite{sabe,bethes}
to a three-dimensional wave equation without relative time, can be absorbed into the momentum dependence of the interaction operator, directly affecting the physical strength 
of the interaction.
To this aim,  we try to study different modelizations of the retardation effects 
that can be used in hadronic constituent  quark interactions.

In the first part of the work,
the use of coordinate space will be emphasized in order to understand in a more intuitive way the contribution of the different terms.
To this aim,
we try to construct a retarded interaction \textit{operator} by using a procedure that relies 
on classical electrodynamics, more precisely, on the well-known Li\'enard-Wiechert (LW) construction, also assuming constant momenta of the interacting particles.

Then, a series expansion of this operator is performed 
in order to calculate
numerically the operator matrix elements.

Finally, we compare our model with the tree-level Feynman diagram calculation, 
showing a substantial physical agreement of  the two approaches.
In this respect, the present work helps to understand  the physical meaning of retardation
in   coordinate space and, in this way,  motivates further calculations
that can be developed directly in momentum space, 
starting from the tree-level Feynman diagram interaction. 

The problem faced in this article presents a high level of complexity, mainly related to
the treatment of time in microscopic systems and to the formulation of 
a relativistic quantum model that could represent an \textit{alternative} 
to standard quantum field theories.
No definitive conclusions will be drawn.

\vskip 0.5 truecm
In more detail, the contents of the paper are organized as follows.
In Subsect. \ref{unret} we recall the construction of a relativistic interaction model 
\textit{without} retardation, commenting on some related developments. 
In Sect. \ref{liwi} the classical derivation, based on the LW construction,
is explained.
Then, in Sect. \ref{quantum} the corresponding quantum operator is constructed 
in different forms, also studying some properties of this operator
 by means of test numerical calculations.
In Subsect. \ref{comments}, some critical comments about the applicability of the model,
in particular for the case of light quarks, will be given.
In Sect. \ref{fey}, we study the relationship between our model and 
the interaction given by the Feynman diagram at tree-level.
Finally, in Subsect. \ref{comgen},   a discussion about the results of this comparison,
in view of possible applications and generalizations, will be performed.

 

Throughout the work, we shall use the standard natural units, 
that is $\hbar=c=1$.

\subsection{Relativistic interaction without retardation}\label{unret}
In this subsection we recall synthetically
that retardation is not strictly necessary to formulate a relativistic interaction model.
This point has been discussed in Ref. \cite{relvar} considering the relativistic (integral) wave equation 
of that model, written in the momentum space.
Physically, an unretarded potential would correspond to make the hypothesis
that the interaction is mediated by
a static field, 
in contrast with the standard mechanism of a bosonic particle interchange.

Here we note that in the coordinate space, one can always define the CM interparticle distance
$r$ in a relativistic invariant form, by using a standard trick \cite{hag}.
Let us consider, in a generic reference frame,  the 4-vector distance between two events:
\begin{equation}\label{rmugen}
r^\mu=(r^0, \vec r)
\end{equation}
where the time component $r^0$ can take \textit{any} form and $\vec r$ is the interparticle
3-vector distance.
We introduce the \textit{projected} 4-vector distance
\begin{equation}\label{rhomu}
\rho^\mu=r^\mu -\left( {\frac {r_\nu P^\nu} {M} } \right) {\frac {P^\mu} {M} } 
\end{equation}
where $P^\mu$ represents the total 4-momentum of the bound system and
$M$ its mass.
In the CM,  $ \rho^\mu$   takes the form
\begin{equation}\label{rhomucm}
\rho^\mu_{_{CM}}= ( 0, \vec r_{_{CM}} )~.
\end{equation}
In this way, in the CM, we get rid of the time component $r^0$ and
the interparticle invariant distance can be written as
\begin{equation}\label{rinv}
r= \sqrt{ - \rho^\mu \rho_\mu}
\end{equation}
where, in the \textit{l. h. s.}, the label \textit{CM} has not been written explicitly.
Note that the last equation defines the variable that is used  in  the
instantaneous potential functions.

As a side remark, we mention the possibility of introducing an interaction given 
by the sum of an unretarded contribution plus a retarded term,
as it is done in the Coulomb gauge in quantum electrodynamics \cite{lali4}.  
We shall not develop this point in the present paper.


\section{The standard LW construction}\label{liwi}
The problem of  retardation has been faced  in the context
of classical electrodynamics
\cite{laliclas,jack},
 highlighting that  the finite speed of light, 
that originates retardation, does not allow to write down an \textit{exact} Hamiltonian
for two (or more) interacting particles.
The retardation effects are taken into account as terms of relativistic corrections
in an expansion of the Hamiltonian in powers of $p/m$.
In this work we try to follow a different method,
writing an (initially) exact expression for the retarded interaction 
not including  the acceleration terms that,
classically, are related  to the radiation emission.
This last effect. in a quantum model,
should be  treated apart, in a different way.

Along the derivation, we bear in mind that
the expression for the interaction  obtained by means of the classical calculation 
must be transformed into a quantum operator for our study 
of hadronic systems.
This last point will be fully developed  in Sect. \ref{quantum}.


%
We start our calculation deriving the expression of the \textit{retarded distance}
between two interacting particles.
Then, we shall use this quantity in the LW retarded interaction.

We recall that, in  quantum mechanics, 
the \textit{standard distance} between two  particles is represented by an Hermitian operator
given by the difference of the position operators of the two particles, 
at the same time.
On the other hand, we want to derive the \textit{retarded distance} operator
whose expression  depends (as it will be shown) on the \textit{standard distance} and on the
\textit{momenta} of the two particles.
We perform the following derivation according to a classical procedure
\cite{laliclas,jack},
with the aim of obtaining, finally, a quantum operator in which the relative time variable
does not explicitly appear.
In a given reference frame, the particle 1,
  at the time $t'$ and
from  the spatial  point $\vec r_1'=\vec r_1(t') $, ``emits" the vector field.
The emission event is defined as $r_1'^\mu=(t', \vec r_1')$.
The field  is ``observed" (\textit{i. e.} ``received") later,
at the time $t>t'$, in the spatial point $\vec r_o$; 
in this way, we can define the observation  event
$r_o^\mu=(t, \vec r_o)$.
At the  time $t$, the position of the particle 1 is $\vec r_1=\vec r_1(t)$, 
that defines the event $r_1^\mu=(t, \vec r_1)$.
We can introduce the \textit{simultaneous} 4-vector distance:
\begin{equation}\label{revdef}
r^\mu=r_o^\mu-r_1^\mu=(0, \vec r)
\end{equation}
where $\vec r$ represents  the \textit{standard vector distance} between
the observation point and the particle 1, at the same time $t$. 
Incidentally,
note that, in the classical theory, $\vec r$ depends on the time $t$ through $\vec r_1$.
For the following developments, recall that, at the time $t$, the velocity of the particle 1 is  $\vec v_1$.

Given  that the speed of the light ($c=1$) is \textit{finite}, the observed field
depends on the  position and velocity of the particle 1 at the  emission time $t'$.
The emission time and position are usually called \textit{retarded} time and position.
Consequently, we can introduce  the \textit{retarded, light-like}, 4-vector distance:
\begin{equation}\label{retadef}
r'^\mu=r_o^\mu-r_1'^\mu
\end{equation}
Considering the time difference
$\Delta t= t-t' > 0$ and taking into account that the field propagates at the
speed of light $c=1$, 
we have:

\begin{equation}\label{del_tc_rr}
r'^{\, 0}= \Delta t=|\vec r'|=r'~,
\end{equation}
that expresses the light-like character of $r'^\mu$,
and
\begin{equation}\label{rvrime}
\vec r' =\vec r_o- \vec r_1'~.
\end{equation}
%
We make the \textit{additional} hypothesis that in the interval between $t'$ and $t$ the particle 1 moves with 
the  constant velocity $\vec v_1$.
This hypothesis amounts to disregard, in the classical theory,  radiation emission.
By using $\Delta t$ of eq. (\ref{del_tc_rr}),
we have for the particle's displacement:
\begin{equation}\label{displacer1}
\vec r_1- \vec r_1'= \vec v_1 \Delta t=     \vec v_1  r'
\end{equation}
that gives, 
for the spatial components of the retarded distance,
the following expression:
\begin{equation}\label{displacerr}
\vec r'=  \vec v_1  r' + \vec r~.
\end{equation}
This equation simply means that the retarded vector distance $\vec r'$ 
is given 
by the sum of the particle's displacement  in the time  interval $\Delta t$
 (first term of the \textit{r. h. s.})
plus the simultaneous spatial vector distance (second term of the \textit{r. h. s.}).
The time component associated to $\vec r'$ is straightforwardly given by eq. (\ref{del_tc_rr}).
%
We can now solve
Eq. (\ref{displacerr})  analytically to find the expression of $ r'$
as a function of $\vec r$ and $\vec v_1$. 
The result is:
\begin{equation}\label{rprime}
r'= \gamma_1^2 \left[ \sqrt{(\vec r \cdot \vec v_1)^2 +{\frac {r^2} {\gamma_1^2} }} 
  +  \vec r \cdot \vec v_1 \right]
	\end{equation}
where 
 $\gamma_1=1/\sqrt{1- \vec v_1^2}$
is the standard Lorentz factor of the particle 1. 

The result of Eq. (\ref{rprime}) can be now used to determine the 4-vector LW potential 
observed at $\vec r_o^\mu $, that is $A^\mu(t,\vec r_o)$.
The general expression for this quantity,
given for the electromagnetic field in Refs. \cite{laliclas,jack},
has the following form:
\begin{equation}\label{lwpot}
A^\mu(t,\vec r_o)= q_1 u^\mu_1 \cdot
{\frac {1}  {d_1}}
\end{equation}
where $ q_1$ represents here the generalized strong charge 
of the particle 1, 
$u_1^\mu$
is the standard particle 4-velocity 
and, finally,  the denominator $d_1$ is given by
\begin{equation}\label{denom0}
d_1=u_1^\nu r'_\nu ~.
\end{equation}
%
Note that $d_1$ of Eq. (\ref{denom0}) depends on 
the \textit{retarded} 4-vector distance  $r'^\mu$.
On the other hand,
due to our hypothesis of constant velocity, it is not necessary to specify that
 $u_1^\nu$ is taken at the time $t'$.

We now elaborate  Eq. (\ref{denom0}) in order to express $d_1$ by means 
of $\vec r$ and of the 3-momentum $\vec p_1$. 
First, 
by using the standard  4-velocity definition, $d_1$ can be written the form:
\begin{equation}\label{denom1}
d_1=\gamma_1 (r'-\vec v_1 \cdot \vec r')~.
\end{equation}
Then, we multiply  Eq. (\ref{displacerr}) by $\vec v_1$ and replace
the expression obtained for $\vec v_1 \cdot \vec r'$ in Eq. (\ref{denom1}).
We have:
\begin{equation}\label{denom2}
d_1={\frac {r'} {\gamma_1}} -  \vec r \cdot \vec u_1~.
\end{equation}
We also replace the expression of $r'$ given by Eq. (\ref{rprime}) in Eq. (\ref{denom2})
and, finally,
recalling that $\vec u_1= \vec p_1/m$, we obtain the searched result:
\begin{equation}\label{denom4}
d_1=\sqrt{ \vec r^2 +  {\left( {\frac { \vec r \cdot   \vec p_1} {m}}\right)}^2 }~. 
\end{equation}
The complete expression of the field takes the form
\begin{equation}\label{lwpotprac}
A^\mu(t,\vec r_o)= q_1 u^\mu_1 \cdot
{\frac {1} {\sqrt{ \vec r^2 +  {\left( {\frac { \vec r \cdot   \vec p_1} {m}}\right)}^2 } }}~.
\end{equation}
%


In order to construct an invariant \textit{interaction} between two particles, we assume that the interacting particle 2, with generalized charge $q_2$, is placed at the observation point,
that is $r_o^\mu= r_2^\mu$.
In this way, instead of Eq. (\ref{revdef}),
we have, \textit{from  now on}, the following definition:  
\begin{equation}\label{r12inst}
r^\mu= r_2^\mu -r_1^\mu =(0,\vec r)= (0,\vec r_2 -\vec r_1)~.
\end{equation}
%
%
Furthermore, we make the hypothesis that the particle 2 moves with 
the constant velocity $\vec v_2$, being $u_2^\mu$
the corresponding 4-velocity.
In this way, the invariant interaction at the time $t$ takes the standard form
\begin{equation}\label{wstart}
W_{12}=q_2 u_2^\mu A_\mu(t,\vec r)
\end{equation}
where $A_\mu(t,\vec r)$ is given by Eq. (\ref{lwpotprac}).

We develop the last expression in the following way.
First, in order to construct the quantum model in Sec. \ref{quantum}, 
we can formally drop the dependence on the (generic) time $t$,
because in the Schr\"odinger representation, the operators do not depend on time.
Furthermore, we specialyze the interaction to a \textit{$q ~ \bar q$ system}.
In this case,
the product of the two strong charges gives
\begin{equation}\label{prodcharge}
q_1 q_2= - {\frac 4 3}  \alpha_s~.
\end{equation}
where $- 4/3$   represents the quark color factor and
$ \alpha_s $ is the strong effective coupling constant.
We obtain:
\begin{equation}\label{w12}
W_{12}=- {\frac 4 3}  \alpha_s u_1^\mu u_2^\nu g_{\mu \nu} \cdot
{\frac {1} {\sqrt{ \vec r^2 +  {\left( {\frac { \vec r \cdot   \vec p_1} {m}}\right)}^2 } }  }
\end{equation}
Now we have to consider the contribution $W_{21}$ that is obtained from Eq. (\ref{w12})
by interchanging the two particles.
Then, we sum the two contributions multiplying by a factor $1/2$ primarily to avoid
double counting.
The result for the total retarded interaction is:
\begin{equation}\label{w}
W_{ret}(\vec r,\vec p_1, \vec p_2)
=- {\frac 4 3}  \alpha_s u_1^\mu u_2^\nu g_{\mu \nu} \cdot
{\frac 1 2}
\left[
{\frac {1} {\sqrt{ \vec r^2 +  {\left( {\frac { \vec r \cdot   \vec p_1} {m}}\right)}^2 } }  }+
{\frac {1} {\sqrt{ \vec r^2 +  {\left( {\frac { \vec r \cdot   \vec p_2} {m}}\right)}^2 } }  }
\right]~.
\end{equation}
In this way we have obtained an expression that is  not only symmetrized with respect the two particles,
but also reconstructs a time-symmetric interaction within the 
stationary  hadronic bound state.
This procedure, that
eliminates the problem of the classical chronological order (where one particle is considered
the ``source" and the  other the ``observer") 
seems highly consistent with the unobservability of the relative time between 
the two particles. 

For practical calculations of bound state properties, 
we introduce, in the CM, the relative momentum 
$\vec p=\vec p_2 = -\vec p_1$.
In this reference frame, the  interaction can be  written 
in the following form:
\begin{equation}\label{lwinteraction}
W_{ret}(\vec r, \vec p)=- {\frac 4 3}  \alpha_s u_1^\mu u_2^\nu g_{\mu \nu}\cdot
V_{ret}(\vec r, \vec p)
\end{equation}
where we have introduced the \textit{spatial function}
\begin{equation}\label{wwork1v}
V_{ret}(\vec r,\vec p) =  
{\frac {1}  {\sqrt{ \vec r^2 + \left( {\frac  {\vec r \cdot\vec p} {m} } \right)^2 }  } }~. 
\end{equation}
%
%
%
%
In the last equation, the denominator of $V_{ret}(\vec r,\vec p)$   represents the \textit{invariant retarded distance} of the two particles.
In Eq. (\ref{lwinteraction}) the 4-velocity product, according to  the 
\textit{classical theory},  gives :
\begin{equation}\label{uiprod}
u_1^\mu u_2^\nu g_{\mu \nu}=1+  {\frac { 2\vec p^2} {m^2}}~
\end{equation}
but, in the quantum model discussed in the next section, the 4-velocities will be
replaced by the Dirac matrices of the two particles.
%

\section{ The quantum operator for bound state calculations}\label{quantum}
In this section, starting from the retarded interaction of Eqs. (\ref{lwinteraction})
and (\ref{wwork1v}), we construct a quantum operator to be used in a reduced two-body Dirac equation.
To this aim, we replace the 4-velocities $u_i^\mu$ of Eq. (\ref{lwinteraction})
with the Dirac matrices of the two particles.
For clarity, we write here the expression that will be elaborated and studied in the 
remainder of the work:
\begin{equation}\label{wwork1}
W_{ret}(\vec r,\vec p)=- {\frac 4 3}  \alpha_s \gamma_1^\mu \gamma_2^\nu g_{\mu \nu}\cdot
V_{ret}(\vec r,\vec p) ~,
\end{equation}
with the spatial factor given by Eq. (\ref{wwork1v}).

Note that in the static case ($\vec p=0$) the retardation
effects disappear.
In the general case,
the retardation effects \textit{reduce} (in absolute value) the strength of the interaction. 

The  expression of  $W_{ret}(\vec r,\vec p) $  of the last equation 
is apparently simple
but cannot be directly ``interpreted" as a quantum
operator  due to the non-Hermitian character of $V_{ret}(\vec r,\vec p)$.    
We have the task to introduce, from that expression 
(obtained by means of a classical procedure), a corresponding Hermitian quantum operator. 
This   operator will have, in any case, a nonlocal form; it means that the retardation effects
involve the wave function of the system over all the space.

The procedure to obtain an Hermitian operator is \textit{not unique}.
In any case, it consists in \textit{choosing an order} of the operators $\vec p$ and $\vec r$
so that an Hermitian operator is finally obtained.
We shall describe in detail the following procedure because  it allows to calculate 
some test 
matrix elements in a simple way, without introducing singular terms.
Another procedure is also incidentally outlined but no specific calculation is performed.
\vskip 0.5 truecm
First, we recall the definition of the following non-Hermitian radial 
momentum operator \cite{sak}:
\begin{equation}\label{pr}
p_r= {\frac 1 r} \vec r\cdot \vec p~.
\end{equation}
For a wave function $\psi(r, \theta, \varphi)=<r, \theta, \varphi  | \psi> $
in spherical coordinates, we have
\begin{equation}\label{prcoord}
<r, \theta, \varphi |p_r| \psi> 
= -i {\frac {\partial} {\partial r} }\psi(r, \theta, \varphi)~.
%
 \end{equation} 
Furthermore, we note that the operator $p_r^\dag p_r$ \textit{is} Hermitian.
For further developments, this operator
can be written, with standard procedures, in the form
\begin{equation}\label{prdagpr}
p_r^\dag p_r=\vec p^2 - {\frac {\vec L^2} {\vec r^2}}~.
\end{equation}
%
For transforming Eq. (\ref{wwork1v})
we  use  the following operator ordering: 
\begin{equation}\label{repl1}
 \left( {\frac {\vec r \cdot \vec p} {r} } \right)^2 \rightarrow
 \vec p \cdot \vec r   {\frac {1} {r^2}}  \vec r \cdot \vec p =
 p_r^\dag p_r~.
\end{equation}
In this way
we can rewrite Eq. (\ref{wwork1})  in the form:
\begin{equation}\label{wwork2}
W_{ret}(\vec r,\vec p)=- {\frac 4 3}  \alpha_s \gamma_1^\mu \gamma_2^\nu g_{\mu \nu}\cdot
{\frac {1}  {r\sqrt{ 1 +  {\frac  {p_r^\dag  p_r} {m^2} }  }  } } ~.
\end{equation}
The previous expression is not yet Hermitian but we can perform a series expansion
in powers of $p_r^\dag p_r/m^2$ and write each term of the series in Hermitian form. 
To this aim we  make use of the following operator reordering:
\begin{equation}\label{rep2}
{\frac 1 r} \left(p_r^\dag p_r \right)^k \rightarrow
 (p_r^\dag)^k {\frac 1 r} (p_r)^k~.
\end{equation}
In consequence, the interaction can be expanded in the following way
\begin{equation}\label{wexp}
W_{ret}(\vec r,\vec p)=- {\frac 4 3}  \alpha_s \gamma_1^\mu \gamma_2^\nu g_{\mu \nu}\cdot
\sum_{k=0}^\infty c_k
\left( {\frac {p_r^\dag} {m} } \right)^k {\frac 1 r} \left( {\frac {p_r} {m} }\right)^k
\end{equation}
where the coefficients $c_k$ are: 
\begin{equation}\label{coefs}
c_k= (-1)^k{\frac {(2k-1) !!} {(2k)!!}     }~.
\end{equation}
We can now study some developments of the previous analysis.
First, we recall that, instead of  the pure Coulombic potential,
it is possible to use a regularized function
$U(r)$ that is introduced in a model in which the quarks are considered as 
extended sources of the interaction field.
In general, we can perform the following formal replacement in Eq. (\ref{wexp}):
\begin{equation}\label{wreg}
{\frac 1 r} \rightarrow  U(r)
\end{equation}
where $U(r)$ indicates a \textit{generic} (pure Coulombic or regularized) spatial potential function.
In particular the author successfully used the following regularized function \cite{rednumb}:
\begin{equation}\label{ureg}
U(r)={\frac 1 r} \text{erf}\left({ \frac {r} {2d}} \right)
\end{equation}
that is obtained considering a Gaussian color charge distribution of the source particles.
Some test calculations performed  with the pure Coulombic interaction and with the regularized one will be discussed below.

Incidentally,
we also show here a different operator ordering that can be used  to  obtain an Hermitian  interaction operator 
$W_{ret}(\vec p, \vec r)$.
Starting  directly  from Eq. (\ref{wwork2}), we define:
\begin{equation}\label{wexpalt}
W_{ret}(\vec r,\vec p) =- {\frac 4 3}  \alpha_s \gamma_1^\mu \gamma_2^\nu g_{\mu \nu}\cdot
\left[ 1+ {\frac {p_r^\dag p_r} {m^2} } \right]^{-1/4}    
U(r)
\left[ 1+ {\frac {p_r^\dag p_r} {m^2} } \right]^{-1/4}  
\end{equation}
where the operator $p_r^\dag p_r$ has been given explicitly in Eq. (\ref{prdagpr}).
Also in Eq. (\ref{wexpalt}) it is possible to take for $U(r)$
 a generic  potential function.
We note that Eq. (\ref{wexpalt}) seems more compact than Eq. (\ref{wexp}).
However, for practical calculations, one has to perform, also in this case,
a series expansion  in powers of $ p_r^\dag p_r/ m^2 $. 
The calculation of the corresponding matrix elements is not 
easy to handle and, for this reason, we do not use the interaction of Eq. (\ref{wexpalt}) in this work.

\vskip 0.5 truecm
We now discuss a test calculation of some matrix elements of the spatial operator 
that appears in Eq. (\ref{wexp}), with the generalization 
of Eqs. (\ref{wreg}) and (\ref{ureg}).
To this aim, we disregard the Dirac matrices and the color coupling,
focusing our attention on
the spatial operator $S$, defined  as:
\begin{equation}\label{sdef}
S= \sum_{k=0}^\infty c_k
\left( {\frac {p_r^\dag} {m} } \right)^k U(r) \left( {\frac {p_r} {m} }\right)^k~.
\end{equation} 
We study the matrix elements
$ < n_b, l | S | n_a, l> $
where $n_b,~ n_a$ represent the energy quantum number of the states and $l$ is
the orbital angular momentum quantum number.
The matrix elements do not depend on the third component of the angular momentum.
For this reason, the corresponding quantum number $m_l$ has been dropped.

For our test calculation we take the standard harmonic oscillator wave functions
in coordinate space given, for example, in Ref. \cite{localred} .
By using the adimensional variable $s=r \bar p$, with the momentum 
scale $\bar p$, 
the matrix elements are written in  the form:
\begin{equation}\label{smatho}
< n_b, l | S | n_a, l>=  \sum_{k=0}^\infty c_k
\left({\frac {\bar p} {m } }\right)^{2k}
\int_0^\infty s^2 ds  \hat R_{n_b,l}^{(k)}(s) 
U \left( {\frac {s} {\bar p} } \right) 
                      \hat R_{n_a,l}^{(k)}(s)~
\end{equation}
where $\hat R_{n,l}^{(k)}(s)$ represents the $k$-th derivative 
with respect to $s$ 
of the adimensional radial functions.
The integrals of the previous expression can be calculated analytically for the
pure Coulombic case. For the regularized $U(r)$ of Eq. (\ref{ureg}) the integrals are 
 performed numerically.
The first term of the series (that is the term with $k=0$) gives  the unretarded interaction
matrix elements.
By considering the calculations of the work \cite{rednumb}, the value of $\bar p$ has been fixed,
indicatively, at the value $\bar p=$ 0.5 GeV;
the value of $d$ in the regularized interaction of Eq. (\ref{ureg}),
at the value $d= 0.15$ fm and the quark mass at $m=1.27$ GeV.
Some results are shown in Table \ref{tab}.
The sum over $k$ has been performed up to $k=8$ with an accuracy of the order of $6 \%$.
The difference betweeen the \textit{unretarded} and \textit{retarded} calculation is
not big but  should be taken into account in more refined models for the charmonium 
and bottomonium spectra. 
\subsection{Comments on the limitations of the model}\label{comments}
Some critical comments about the general applicability  of the model are necessary.
The model has been developed with the aim to improve the description of
\textit{heavy} $q ~ \bar q$ systems.
For the case of light quark mesons, the non-perturbative character of confinement and 
the relativistic speeds of the quarks are likely to invalidate the treatment,
because, for these states, the mean values of the powers of $p/m$ become large quantities.
In particular, the expansion in powers of $ \bar p /m $ 
of Eq. (\ref{smatho})  is likely to diverge. 
Furthermore,  different  operator orderings
(whose effects appear in higher-order terms of $p/m$) can give significantly different
numerical predictions, showing the arbitrariness of the ordening choice.
Finally, the constant quark velocity assumption, neglecting the radiation effects
(classicaly related to acceleration) may be questionable for non-perturbative, 
highly relativistic  bound systems.
For all these reasons, we conclude that the LW construction
should be merely considered  as a \textit{correction}
and  is probably insufficient to treat the dynamics of light quark 
mesons.

\begin{table*} 
\caption{
Test calculation of some retarded and unretarded matrix elements of the spatial 
operator $S$ of Eq. (\ref{smatho} ). 
In the first three columns we report
$n_a$, $n_b$ and $l$ that represent the energy quantum number of the \textit{ket} state, 
the energy quantum number of the \textit{bra} state and the angular momentum quantum number,
respectively.
In the other columns we report the values of the matrix elements in the 
\textit{unretarded} case (subscript \textit{unr}) and in the 
\textit{retarded} case (subscript \textit{ret}).
The calculations have been performed for a pure Coulombic interaction
(superscript \textit{C}) and for the regularized interaction (superscript \textit{R})
of Eq. (\ref{ureg}).
All the results are in GeV. Further details are given in the text.
}
\begin{center}
\begin{tabular}{lllllllll}
\hline \\
$n_a$  & $n_b$   & $l$ &  $S_{unr}^C$ & $S_{ret}^C$ & $S_{unr}^{R}$ &$S_{ret}^{R}$  & \\     
\hline \\   
$0   $ &  $0 $ & $0$  & $0.564  $ & $0.525 $  &  $0.449  $ & $0.412 $ & \\
$0   $ &  $2$  & $0$  & $0.230  $ & $0.183 $  &  $0.116  $ & $0.0743 $ &   \\
$1   $ &  $1$  & $1$  & $0.376  $ & $0.354 $  &  $0.354  $ & $0.335  $ &   \\
$2   $ &  $2$  & $0$  & $0.470  $ & $0.368 $  &  $0.350  $ & $0.261  $ &   \\
$2   $ &  $2$  & $2$  & $0.301  $ & $0.280 $  &  $0.295  $ & $0.277  $ &   \\
$3   $ &  $1$  & $1$  & $0.119  $ & $0.103 $  &  $0.093  $ & $0.0786 $ &   \\
$3   $ &  $3$  & $1$  & $0.338  $ & $0.285 $  &  $0.306  $ & $0.251 $ &   \\
$3   $ &  $3$  & $3$  & $0.258  $ & $0.245 $  &  $0.256  $ & $0.243 $ &   \\
\hline \\
\end{tabular}
\end{center}

\label{tab}
\end{table*}

\section{A comparison with the Feynman model}\label{fey}
In this section we discuss a point of conceptual interest, that is the comparison 
of our model of interaction (developed in the coordinate space)
with the results, at tree-level, of the Feynman diagram theory
that are obtained in the momentum space.
The following  analysis can also represent the starting point for constructing a momentum space 
model of the quark interaction.
Some comments and a possible generalization of this model will be presented 
in Subsect. \ref{comgen}.
 

We develop the calculations of this section by using
the retarded interaction of Eq. (\ref{wwork1}) 
with  $V_{ret}(\vec r,\vec p)$ of Eq. (\ref{wwork1v}),
that is we consider point-like interacting quarks.
The generalization to the case of non-point-like quarks will be discussed 
in Subsect. \ref{comgen}.

For clarity,  we previously recall  some standard procedures
that are used for the \textit{unretarded}
(or \textit{static}) interaction that is obtained 
by setting $\vec p=0$ in Eq. (\ref{wwork1v}).
In this case,
we have a pure Coulombic  interaction operator, in the form
\begin{equation}\label{wunr}
W_{unr}(\vec r)=- {\frac 4 3}  \alpha_s \gamma_1^\mu \gamma_2^\nu g_{\mu \nu}\cdot
{\frac {1}  { r } }~.
\end{equation}
In the momentum space, we have
\begin{equation}\label{wunrq}
\hat W_{unr}(\vec q)= <\vec p_b | W_{unr}(\vec r) |\vec p_a>
\end{equation}
where $\vec q=\vec p_a -\vec p_b$ is the interaction 3-momentum transfer.
We introduce for the present work the definition \textit{interaction amplitude}
for the momentum space matrix elements of the interaction operator.
Specifically, the unretarded interaction amplitude
$\hat W_{unr}(\vec q)$ is obtained by means of a standard Fourier transform:
\begin{equation}\label{ftstand}
\begin{split}
\hat W_{unr}(\vec q) &={\frac {1} {(2 \pi) ^3} } \int d^3 r \exp( i \vec q \cdot \vec r)
     W_{unr}(\vec r)\\
                &= - {\frac 4 3}  \alpha_s \gamma_1^\mu \gamma_2^\nu g_{\mu \nu}\cdot
{\frac {1} {2 \pi^2} } \cdot {\frac {1} {\vec q^2}}
\end{split}
\end{equation}
where the expression shown in the second line 
is obtained by means of standard handlings.
%
We recall that, in general,  the dependence 
of  $ \hat W_{unr}(\vec q) $
on $\vec q$ \textit{only},
corresponds to the local, unretarded character of   $W_{unr}(\vec r)$.

On the other hand, in the Feynman theory, by using the tree-level  diagram \cite{lali4}, 
one has for the interaction amplitude, an expression of the following kind:
\begin{equation}\label{feyn}
\hat W_{Fey}(\vec p_b, \vec p_a) 
                = - {\frac 4 3}  \alpha_s \gamma_1^\mu \gamma_2^\nu g_{\mu \nu}\cdot
{\frac {1} {2 \pi^2} } \cdot {\frac {1} {\vec q^2 -(\Delta \epsilon)^2}}\cdot
N(p_b,p_a)
\end{equation} 
where the energy difference $\Delta \epsilon$, 
represents the time component  $q^0$ of the on-shell 4-momentum transfer.
It is standardly defined as:
\begin{equation}\label{deltaeps}
q^0=\Delta \epsilon= \epsilon(\vec p_a) - \epsilon(\vec p_b)
\end{equation}
Physically, it gives rise to the non-local and retarded character of the interaction,
whereas it is neglected in unretarded calculations.
We also note that the invariant (positive) squared momentum transfer
\begin{equation}\label{q2}
Q^2=- q^\mu q_\mu= \vec q^2 -(\Delta \epsilon)^2
\end{equation}
standardly appears in the denominator of Eq. (\ref{feyn}).
We have also introduced the relativistic factor  $N(p_b,p_a)$
that will be discussed in the following.
We anticipate that this factor reduces to one in the nonrelativistic limit. 
%

Analogously to Eq. (\ref{wunrq}), it is possible to write \textit{formally}
the interaction amplitude
$\hat W_{Fey}(\vec p_b, \vec p_a)$ as a matrix element between two momentum states:
\begin{equation}\label{wform}
\hat W_{Fey}(\vec p_b, \vec p_a) = 
<\vec p_b | W_{Fey} |\vec p_a>
\end{equation}
but, in this case, the explicit expression of the nonlocal operator $W_{Fey}$ 
is not directly given by the theory.

 
In order to compare the Feynman interaction amplitude  with our LW retardation model, 
it is also useful to write
the on-shell energy difference in the following way
\begin{equation}\label{derewr}
\Delta \epsilon= - ~{\frac {\vec q \cdot (\vec p_a+\vec p_b)} 
{\epsilon(\vec p_b) + \epsilon(\vec p_a) } }~.
\end{equation}
In consequence, the Feynman interaction amplitude of Eq. (\ref{feyn})
can be  rewritten in the completely equivalent  form:

\begin{equation}\label{feynrew}
\hat W_{Fey}(\vec p_b, \vec p_a) 
                = - {\frac 4 3}  \alpha_s \gamma_1^\mu \gamma_2^\nu g_{\mu \nu}\cdot
{\frac {1} {2 \pi^2} } \cdot {\frac {1} 
{\vec q^2 -
\left[  {\frac {\vec q \cdot (\vec p_a+\vec p_b)} 
{\epsilon(\vec p_b) + \epsilon(\vec p_a) } }  \right]^2}} 
\cdot N(p_b,p_a).
\end{equation} 
\vskip 0.5 truecm
We  now turn our attention to the retarded the LW   interaction operator of Eq. (\ref{wwork1}), 
with $V_{ret}(\vec r, \vec p)$
of Eq. (\ref{wwork1v}),
and introduce the corresponding interaction amplitude 
\begin{equation}\label{vhatret}
\hat W_{ret}(\vec q, \vec p) =<\vec p_b | W_{ret}(\vec r, \vec p) |\vec p_a>~.
\end{equation} 
Analogously to Eq. (\ref{ftstand}), the previous expression can be written by means of
the following integral:
\begin{equation}\label{ftret}
\hat W_{ret}(\vec q, \vec p) ={\frac {1} {(2 \pi) ^3} } \int d^3 r \exp( i \vec q 
\cdot \vec r) W_{ret}(\vec r, \vec p)
\end{equation}

%
To calculate explicitly the previous quantity 
we \textit{generalize}  the Fourier transform  of Eq. (\ref{ftstand}).
In this calculation  the momentum $\vec p$ will be temporarily  treated as a constant,
disregarding, for the moment, 
the operator ordering.
The procedure of this  generalized calculation 
can be schematized as follows.

Instead of integrating with respect to $\vec r$,
we introduce $\vec \eta$ as new  spatial integration variable:
\begin{equation}\label{eta}
\vec \eta=\vec r +{\frac {(\vec r \cdot  \vec p )\vec p } {m \left[m+\epsilon(
\vec p) \right]}}~;
\end{equation}
$\vec \eta$ satisfies the following identity:
\begin{equation}\label{etasq}
\vec \eta ^2=   \vec r^2 + \left( {\frac  {\vec r \cdot\vec p} {m} } \right)^2
 ~. 
\end{equation}
By means of this expression we can write  the retarded spatial function of Eq. (\ref{wwork1v})
in the form:
\begin{equation}\label{vreteta}
V_{ret}(\vec r, \vec p)= {\frac 1 \eta} ~.
\end{equation}
We also need to express $\vec r$ as a function of $\vec \eta$. One has:
\begin{equation}\label{reta}
\vec r=\vec \eta - \left( {\frac {\vec \eta \cdot \vec p} {m} }\right)
{\frac {\vec p} {m} }
{\frac {m^2} {\left[m + \epsilon(\vec p) \right] \epsilon (\vec p) } }~.
\end{equation}
From this equation we obtain the
 Jacobian determinant $J$ for the variable change of Eq. (\ref{eta}):
\begin{equation}\label{jacob}
J={\frac {\partial (\vec r) } {\partial (\vec \eta)} }= {\frac {m} {\epsilon(
\vec p)} }~.
\end{equation}
Then  we introduce the constant vector $\vec \xi$:
\begin{equation}\label{xi}
\vec \xi=\vec q - {\frac {\left(  {\vec q \cdot \vec p}  \right) \vec p}
                  {\left[m + \epsilon(\vec p) \right] \epsilon (\vec p) } }
\end{equation}
that  satisfies the two following identities:
\begin{equation}\label{id1}
\vec q \cdot \vec r = \vec \xi \cdot \vec \eta~,
\end{equation}
\begin{equation}\label{id2}
\vec \xi^2 = q^2 - \left[ {\frac {\vec q \cdot \vec p} {\epsilon(\vec p)} 
} \right]^2~.
\end{equation}
In Eq. (\ref{ftret}), by means of Eq. (\ref{id1}), we can replace 
 $\vec q \cdot \vec r$ with $\vec \xi \cdot \vec \eta$.
In this way, in the final result of the integration over $\vec \eta$,
the  factor $1/\vec q^2$ of the unretarded calculation is replaced
by $ 1/ \vec  \xi ^2$.
By using Eq. (\ref{id2}), one obtains the final result:
\begin{equation}\label{wretfin}
\hat W_{ret}(\vec q, \vec p) = - {\frac 4 3}  \alpha_s \gamma_1^\mu \gamma_2^\nu g_{\mu \nu} \cdot {\frac {1} {2 \pi^2} } \cdot 
{\frac {1} {\vec q^2 -
\left[  {\frac {\vec q \cdot \vec p} 
{\epsilon(\vec p)  } }  \right]^2}}
\cdot {\frac {m} {\epsilon(\vec p)}}
~.
\end{equation} 
We can now put this result of our LW model of retardation 
(transformed to momentum space) 
in the same form as 
the Feynman interaction amplitude, given 
in Eq. (\ref{feynrew}).
%
%
To this aim,
in our result of Eq. (\ref{wretfin}), we  make the following replacements:
\begin{equation}\label{replacements}
\begin{split}
\vec p &\rightarrow {\frac 1 2} \left( \vec p_b + \vec p_a \right) ~,\\
\epsilon (\vec p) &\rightarrow {\frac 1 2} 
 \left[ \epsilon(\vec p_b) +\epsilon(\vec p_a) \right]~.
\end{split}
\end{equation}
With these replacements, one recovers the same expression given by the 
Feynman theory of Eq. (\ref{feynrew}).
In particular, 
from the  factor $m/\epsilon(\vec p)$ 
(that represents the Jacobian determinant of Eq. (\ref{jacob}) )
one obtains  for the relativistic factor $N(p_b,p_a)$  the following expression: 
\begin{equation}\label{nfey}
N(p_b,p_a)={\frac {2 m} { \epsilon(\vec p_b) +\epsilon(\vec p_a)} }~.
\end{equation}
All this procedure shows a substantial physical agreement
of the two models.
Further discussions and comments will be given in the following Subsect. \ref{comgen}.




 
\subsection{ Comments on the model and possible generalizations}\label{comgen}
We analyze here the physical content of the obtained results and discuss some possible
applications and generalizations.

Starting from  the LW interaction operator of our model, we have obtained, 
in Eq. (\ref{wretfin}), the corresponding interaction amplitude.
By using, as \textit{prescriptions},  the replacements of Eq. (\ref{replacements}),
we have reconstructed the standard interaction amplitude of the tree-level
Feynman theory.
In consequence, we conclude tentatively that the latter theory has the same physical content
of our LW retardation model.

The Feynman theory gives a nonlocal interaction amplitude from which,
at the moment,
a standard Hermitian operator (depending on $\vec r$ and $\vec p$)
cannot be explicitly derived. 
However, one can use \textit{directly} the Feynman interaction amplitude to calculate,
in the momentum space,
the  bound state matrix elements.
In this case, it is necessary to
use  the momentum space wave functions of the bound states;    
the price to pay, with respect to standard coordinate space calculations, 
is that numerical integrals of higher dimensionality must be performed.
 
On the other hand, our coordinate space LW retarded interaction allows
to understand more intuitively the role of retardation in bound systems.
However, to obtain an Hermitian expression, an operator ordering,
not univocally defined, must be introduced.
Furthermore, in any case, for calculating the bound state matrix elements
an expansion in powers of $p/m$ (not convergent for light quarks)
must be performed.

We conclude that, for introducing retardation in hadronic models, 
the Feynman interaction amplitude seems preferable to the LW coordinate
interaction that, on the other hand, can be used to  check  
more sophisticated momentum space models. 

Some words of caution are needed about the relativistic factor $N(p_b,p_a)$
explicitly given in  Eq. (\ref{nfey}).
That expression is the result of the replacements given by Eq. (\ref{wretfin})
that, as stated before, represent a prescription to reconstruct 
the Feynman interaction amplitude. 
Other forms could be hypothesized for that quantity; 
for example, one could ``symmetrize" separately, in the following  way,
the Jacobian determinant of Eq. (\ref{jacob}), obtaining
\begin{equation}\label{nfeyhyp}
N'(p_b,p_a)={\frac m  2 } 
 \left[ {\frac {1} {\epsilon(\vec p_b)} } +
        {\frac {1} {\epsilon(\vec p_a)} }  
 \right]~.
\end{equation}
Different forms of  $N(p_b,p_a)$ can give significantly different numerical results.
This point should be carefully considered when the theoretical models 
are developed and compared
with the experimental data of the hadronic spectra.

Another point of interest is that for the phenomenological quark interaction
it is usually necessary to introduce  vertex form factors. 
In the unretarded models, one has a color (non-point-like) charge distribution
of the quarks
that gives rise to a \textit{regularized} quark interaction \cite{chromomds,rednumb}
that only depends on $r$.
In the Feynman model, one can multiply the interaction amplitude of Eq. (\ref{feyn})
by  two vertex form factors that depend on the \textit{invariant} (squared)
momentum transfer $Q^2$ of Eq. (\ref{q2}), taking directly into account 
the retardation contributions given by 
$q^0=\Delta \epsilon$, defined in Eq. (\ref{deltaeps}).

Furthermore, we recall that also the strong coupling constant $\alpha_s$
really is a \textit{running}
coupling constant $\alpha_s( Q^2)$ \cite{explor}
whose dependence on $ Q^2$ must be correctly taken into account.

All these additional dependences on $Q^2$ can be straightforwardly included in
a momentum space calculation. 

In summary, a momentum space calculation based on the Feynman interaction amplitude 
 of Eq. (\ref{feyn}), with the form factors and the $\alpha_s( Q^2)$,
should be able to treat retardation and relativistic effects 
that, otherwise, would be  partially ignored in coordinate space models.



\end{document}